\pdfoutput=1
\documentclass[12pt]{article}

\font\grande=cmr9.5 scaled \magstep4
\font\medio=cmr9.5 scaled \magstep2
\outer\def\beginsection#1\par{\medbreak\bigskip
      \message{#1}\leftline{\bf#1}\nobreak\medskip
\vskip-\parskip
      \noindent}
\usepackage{graphicx} 
\begin{document}
\bibliographystyle {unsrt}

\titlepage

\begin{flushright}
CERN-TH-2016-195
\end{flushright}

\vspace{15mm}
\begin{center}
{\grande Hypermagnetic knots and gravitational radiation}\\
\vspace{5mm}
{\grande at intermediate frequencies}\\
\vspace{15mm}
 Massimo Giovannini 
 \footnote{Electronic address: massimo.giovannini@cern.ch} \\
\vspace{0.5cm}
{{\sl Department of Physics, Theory Division, CERN, 1211 Geneva 23, Switzerland }}\\
\vspace{1cm}
{{\sl INFN, Section of Milan-Bicocca, 20126 Milan, Italy}}
\vspace*{1cm}

\end{center}

\centerline{\medio  Abstract}
\vskip 0.5cm
The maximally gyrotropic configurations of the hypermagnetic field at the electroweak epoch 
can induce a stochastic background of relic gravitational waves with comoving frequencies ranging from the 
$\mu$Hz to the kHz.  Using two complementary approaches we construct a physical template family for the emission of the gravitational 
radiation produced by the hypermagnetic knots.  The current constraints and the presumed sensitivities 
of the advanced wide-band interferometers (both terrestrial and space-borne) are combined to infer that the 
lack of observations at intermediate frequencies may invalidate the premise of baryogenesis 
models based (directly or indirectly) on the presence of gyrotropic configurations of the 
hypermagnetic field at the electroweak epoch. Over the intermediate frequency range
the spectral energy density of the gravitational waves emitted by the hypermagnetic knots at the electroweak 
scale can exceed the inflationary signal even by nine orders of magnitude without affecting the standard 
bounds applicable on the stochastic backgrounds of gravitational radiation. The signal of hypermagnetic knots 
can be disambiguated, at least in principle, since the the produced gravitational waves are polarized.
\noindent

\newpage

\vspace{5mm}
\renewcommand{\theequation}{1.\arabic{equation}}
\setcounter{equation}{0}
\section{Introduction} 
\label{sec1}

The equations of the tensor modes of the geometry are not invariant under a Weyl rescaling 
of the four-dimensional metric \cite{one}, as suggested by Grishchuk well before the early formulations 
of conventional inflationary models. This means that, during inflation, the evolution of the space-time curvature induces 
a stochastic background of relic gravitons \cite{one} with a spectral energy 
density extending today from frequencies ${\mathcal O}(\mathrm{aHz})$ (i.e. $1 \,\mathrm{aHz} = 10^{-18} \mathrm{Hz}$) up to frequencies ${\mathcal O}(\mathrm{GHz})$ (i.e. $1\, \mathrm{GHz} = 10^{9} \mathrm{Hz}$).  
In what follows the cosmic background of relic gravitons shall be described in 
terms of the spectral energy density $\Omega_{gw}(\nu,\tau_{0})$, i.e. the energy density of the relic 
gravitational waves (measured in units of the critical energy density) per logarithmic interval of frequency and at the present (conformal) time $\tau_{0}$. 
The description of the energetic content of a stochastic background of gravitational radiation in terms of $\Omega_{gw}(\nu,\tau_{0})$ goes back to the seminal contributions of Refs. \cite{one} and has been consistently used since then to characterize the cosmic graviton background. The astrophysical sources of gravitational radiation, on the contrary, are often described in terms of the spectral amplitude (measured in units of $1/\mathrm{Hz} = \mathrm{sec}$) or even in terms of the dimensionless strain.  The mutual relation between these three quantities can be found, for instance, in  Ref. \cite{thorne} as well as in various more recent  references\footnote{For the sake of conciseness the dependence of $\Omega_{gw}$ upon $\tau_{0}$ will be omitted when not strictly necessary. In this matter we follow the notations and the approach of the last paper of Ref. \cite{onea}.}. To clarify the results obtained here, it seems both useful and relevant to discuss, in some depth, the 
relation between the spectral energy density (i.e. $\Omega_{gw}(\nu,\tau_{0})$), the spectral amplitude (i.e. $S_{h}(\nu,\tau_{0})$ in what follows)
and the dimensionless strain amplitude (i.e. $h_{c}(\nu,\tau_{0})$ in what follows). 

Given the Fourier amplitudes of the tensor modes of the geometry (see, e.g. Eqs. (\ref{TT}) and (\ref{pol1})--(\ref{pol2}))
the power spectrum ${\mathcal P}_{T}(k,\tau)$ appears in the correlation function of the Fourier amplitudes 
\begin{equation}
\langle h_{ij}(\vec{k},\tau) h_{mn}(\vec{p}, \tau) \rangle = \frac{2 \pi^2}{k^3} {\mathcal S}_{ijmn} P_{T}(k,\tau) \,\delta^{(3)}(\vec{k}+\vec{p}), \qquad {\mathcal S}_{ijmn}  = \frac{1}{4}\sum_{\lambda} e^{(\lambda)}_{ij} \,e^{(\lambda)}_{mn},
\label{PT}
\end{equation}
where the result of the sum over the polarizations is given in Eq. (\ref{pol2}); in Eq. (\ref{PT}) $k$ denotes the comoving wavenumber related 
to the comoving frequency in natural units as   $\nu=k/(2\pi)$. At the present time $\tau_{0}$ the tensor power spectrum, the spectral amplitude $S_{h}(\nu,\tau_{0})$ (measured in units of $\mathrm{Hz}^{-1}$)  and the dimensionless strain amplitude $h_{c}(\nu,\tau_{0})$ are related in the 
following manner:
\begin{equation}
{\mathcal P}_{T}(\nu,\tau_{0})= 4 \nu S_{h}(\nu,\tau_{0}),\qquad 
\Omega_{gw}(k,\tau_{0}) = \frac{k^2}{12 {\mathcal H}^2} {\mathcal P}_{T}(k,\tau_{0}).
\label{def1}
\end{equation}
In Eq. (\ref{def1}), as we shall reinstate in the forthcoming sections, ${\mathcal H} = a H$ where $a$ is the scale factor and  $H$ is the Hubble rate. 
As a consequence of Eqs. (\ref{PT}) and (\ref{def1}) the following chain of equalities can be easily derived:
\begin{equation}
{\mathcal S}_{h}(\nu,\tau_{0}) = \frac{3 a_{0}^2 H_{0}^2}{4 \pi^2 \nu^3} \Omega_{gw}(\nu,\tau_{0})=   7.981\times 10^{-43} \,\,\biggl(\frac{100\,\mathrm{Hz}}{\nu}\biggr)^3 \,\, h_{0}^2 \,\Omega_{gw}(\nu,\tau_{0})\,\, \mathrm{Hz}^{-1},
\label{A}
\end{equation}
where $H_{0} = 100\, h_{0}\, (\mathrm{km}/\mathrm{sec}) \mathrm{Mpc}^{-1}$ is the present value of the Hubble rate and  $h_{0} = {\mathcal O}(0.7)$ 
its indetermination which appears as one of the fir parameters of the concordance lore \cite{onea}.
The dimensionless strain amplitude 
obeys $h_{c}^2(\nu,\tau_{0}) =2 \nu S_{h}(\nu,\tau_{0})$ so that  $h_{c}(\nu,\tau_{0})$ becomes explicitly:
\begin{equation}
h_{c}(\nu,\tau_{0}) = 1.263 \times 10^{-20} \biggl(\frac{100 \,\, \mathrm{Hz}}{\nu}\biggr) \, \sqrt{h_{0}^2 \,\Omega_{gw}(\nu,\tau_{0})}.
\label{B}
\end{equation} 
Since $\Omega_{gw}(\nu,\tau_{0})$ is the quotient between the energy density 
of the gravitational waves and the critical energy density (i.e. $\rho_{crit} = 3 H_{0}^2/(8\pi G)$ where, again, $H_{0}$ is the present Hubble rate) it is clear that in $h_{0}^2 \Omega_{gw}(\nu,\tau_{0})$ the dependence on $h_{0}^2$ simplifies between the numerator and the denominator. It is therefore  practical to describe the stochastic backgrounds of gravitational radiation in terms of 
$h_{0}^2 \Omega_{gw}(\nu,\tau_{0})$ which does not depend explicitly upon $h_{0}$.

The normalization of the spectral energy density in critical units for frequencies of the order of the aHz is fixed by the tensor to scalar ratio (conventionally denoted by $r_{T}$). Between the aHz and ${\mathcal O}(100\, \mathrm{aHz})$ a characteristic spectral signal is produced by those modes that exited the Hubble radius during inflation and reentered after the beginning of the matter epoch. Since the analysis of the temperature and polarization anisotropies of the Cosmic Microwave Background \cite{onea} demands that $r_{T} <0.1$,  the spectral energy density computed in conventional inflationary models is bounded from above at all frequencies and, in particular, at intermediate frequencies  $\mu\mathrm{Hz} <\nu <10 \, \mathrm{kHz}$ where  $h_{0}^2 \Omega_{gw}(\nu,\tau_{0}) < 1.6\times 10^{-17}$ (see, in particular, the last paper of Ref. \cite{onea}).
To draw a comparison, the recent detection of gravitational waves reported by the LIGO/Virgo collaboration 
corresponds to a dimensionless strain amplitude $h_{c} = {\mathcal O}(10^{-21})$ which is between eight and nine 
orders of magnitude larger than the stochastic background produced by the conventional inflationary models predicting, from Eq. (\ref{B}),
$h_{c} = {\mathcal O}(10^{-29})$. 

The purpose of the present analysis is to derive a template family for the emission of gravitational waves from maximally gyrotropic configurations of the hypermagnetic field between the $\mu$Hz and  the kHz. The spectral energy density of the gravitational waves obtained in this discussion exceeds the purported inflationary signal and will also be compared with the foreseen sensitivities of wide-band interferometers (both terrestrial and space-borne). In particular, as it will be clear from the phenomenological discussion, the spectral energy density of these peculiar sources can even be $h_{0}^2 \Omega_{gw} = {\mathcal O}(10^{-8})$ for the intermediate frequency range $\mu\mathrm{Hz} < \nu< \mathrm{kHz}$. Assuming, for the sake of concreteness, that $\nu = {\mathcal O}(0.1)$ kHz we have that $h_{c}$ could even be as large as $10^{-24}$, that is to say five orders of magnitude larger than the inflationary signal in the same frequency range. 

The layout of this topical paper is the following. In section \ref{sec2} we shall briefly discuss the connections between 
the hypermagnetic knots and the electroweak physics. The gravitational waves emitted either by a single configuration or by a stochastic collection of hypermagnetic knots will be computed in section \ref{sec3}. Section \ref{sec4} contains the phenomenological considerations. The concluding remarks are collected in section \ref{sec5}.

\renewcommand{\theequation}{2.\arabic{equation}}
\setcounter{equation}{0}
\section{Hypermagnetic knots and electroweak physics} 
\label{sec2}

When the mass of the Higgs boson is ${\mathcal O}(125 \,\mathrm{GeV})$ \cite{oneb} the phase diagram of the electroweak theory 
can only be scrutinized with non-perturbative methods. Lattice simulations \cite{two} suggest the presence of a cross-over regime which 
is conceptually similar to the behaviour experienced by ordinary chemical compounds above their triple point. In this regime the collision 
of bubbles of the new phase cannot happen but gravitational waves can still be emitted provided the electroweak plasma hosts maximally gyrotropic 
configurations of the hypermagnetic field (dubbed {\em hypermagnetic knots} in Ref. \cite{three}) for typical temperatures 
$T_{ew} = {\mathcal O}(100\, \mathrm{GeV})$. Inside the electroweak particle horizon, hypermagnetic knots (HK in what follows) 
can be pictured as a collection of flux tubes (closed because of transversality) but characterized by a non-vanishing 
gyrotropy (i.e. $\vec{B} \cdot \vec{\nabla} \times \vec{B}$ where $\vec{B}$ will denote throughout the comoving hypermagnetic field). The dynamical production of HK and Chern-Simons waves suggested in the past a viable mechanism 
for the generation of the baryon asymmetry of the Universe \cite{three} (see also \cite{four}). While this interesting possibility will be swiftly 
reviewed at the end of this section, as previously pointed out \cite{three} the minimal comoving frequency of the gravitational waves potentially emitted by the HK is: 
\begin{equation}
\nu_{ew} = 25.03  \biggl(\frac{N_{eff}}{106.75}\biggr)^{1/4} \biggl(\frac{T_{ew}}{10^{2}\, \mathrm{GeV}} \biggr) \, \mu\mathrm{Hz},
\label{first}
\end{equation}
where  $N_{eff}$ the effective number of relativistic degrees of freedom at $T_{ew}$. While the smallest frequency of the 
emission is given by Eq. (\ref{first}), the equations of anomalous magnetohydrodynamics \cite{three,five} govern the 
dynamics of the HK and determine, ultimately, the largest frequency of the spectrum.

The non-screened vector modes of the electroweak plasma correspond to the hypercharge field which has a chiral coupling to fermions. The axial currents may arise either as a finite density effect (implying a non-trivial evolution of the chemical potential), or they can be associated with the presence of an axion field. Anomalous magnetohydrodynamics (AMHD) aims at describing the dynamical evolution of the gauge fields in a plasma containing both vector and axial-vector currents: while the axial currents are not conserved because of the triangle anomaly, the vector currents are responsible for the Ohmic dissipation\footnote{The AMHD equations differ from the ones where only chiral currents are present \cite{fivea} at finite fermionic density. They are sometimes presented as the theoretical rationale for the chiral magnetic effect \cite{seven} originally discussed at the electroweak epoch \cite{sevena} and today studied in  the collisions of heavy ions. The name AMHD is technically preferable and goes back to Ref. \cite{sevena} where the chiral magnetic effect was originally discussed by generalizing the analysis of finite density effects \cite{fivea} to the case of finite conductivity.}. Even if AMHD admits a generally covariant formulation (see, in this respect, the last paper of Ref. \cite{five}), for typical wavelengths smaller than the particle horizon at the electroweak epoch\footnote{This means that $k/{\mathcal H}_{ew} = k \tau_{ew}  >1$. Recall that ${\mathcal H}_{ew} = a_{ew} H_{ew}$; $H_{ew}$ is the Hubble rate at the electroweak epoch and $a_{ew}$ is the scale factor.},  the relevant subset of the AMHD equations for the (comoving) hypermagnetic field and for the vorticity $\vec{\omega}$ can be written as \cite{five}:
\begin{eqnarray}
\partial_{\tau} \vec{B} &=& \vec{\nabla}\times(\vec{v} \times \vec{H}) + \frac{\nabla^2 \vec{B}}{ 4 \pi \sigma} + \frac{\vec{\nabla} \times (g_{\omega} \vec{\omega})}{4 \pi \sigma} - 
\frac{\vec{\nabla} \times(g_{B} \vec{B})}{4 \pi \sigma},
\label{seca}\\
\partial_{\tau} \vec{\omega} &=& \vec{\nabla}\times(\vec{v} \times \vec{\omega})  + \frac{\vec{\nabla} \times(\vec{J} \times\vec{B})}{a^4 (\rho + p)} +  \frac{\eta}{a^4(\rho + p)}\nabla^2 \vec{\omega}, 
\label{secb}
\end{eqnarray}
where  $\eta$ is the shear viscosity and  $\sigma$ is the comoving conductivity of the electroweak plasma\footnote{Equations (\ref{seca}) and (\ref{secb}) hold in the case of a conformally flat geometry of Friedmann-Robertson-Walker (FRW) type. Obviously in the radiation case $a^4 (\rho + p) = 4 a^{4} \rho/3=\mathrm{constant}$.}. Equations (\ref{seca}) and (\ref{secb}) hold when the two-fluid effects can be neglected in the slow branch of the AMHD spectrum (see, in particular, the first two papers of Ref. \cite{five}).  
In Eq. (\ref{seca}) $g_{B}$ and $g_{\omega}$ denote, in a concise notation, the coefficients of the magnetic and the vortical currents \cite{three,five}. There are specific situations where the chemical potential and the axion feld can be simultaneously present (see, respectively, the second paper of \cite{three} and the last paper of \cite{five}) both contributing to $g_{B}$ (and possibly to $g_{\omega}$). However, in spite of the potential richness of the spectrum of AMHD (which has been explored elsewhere) for the present ends the most relevant observation is that the perfectly conducting limit suppresses the anomalous contributions.

According to Eq. (\ref{seca}) when the conductivity is very large  the magnetic and the vortical currents are suppressed in comparison with the remaining term (which is the electroweak analog of the standard dynamo contribution). In a shorthand notation we then have that $\partial_{\tau} \vec{B} = \vec{\nabla}\times(\vec{v} \times \vec{B}) + {\mathcal O}(g_{B}/\sigma)$. Defining the vector potential in the Coulomb gauge, Eq. (\ref{seca}) becomes, up to small corrections due to the conductivity, $\partial_{\tau} \vec{A}= \vec{v} \times (\vec{\nabla}\times\vec{A})$. In the highly conducting limit, thanks to classic analyses\footnote{We remind here the results due to Fermi and Chandrasekhar (in connection with the stability of galactic arms) and, a little later,  the analyses of  Chandrasekhar Kendall and Woltjer \cite{six}.}  the magnetic energy density shall then be minimized in a fiducial volume $V$ under the assumption of constant magnetic helicity by introducing the Lagrange multiplier $\zeta$. The configurations minimizing the functional 
\begin{equation}
{\mathcal G} =\int_{V} d^{3} x\{ |\vec{\nabla} \times \vec{A}|^2 - \zeta \vec{A} \cdot (\vec{\nabla}\times\vec{A}) \},
\label{secd}
\end{equation}
are the Beltrami fields satisfying $\vec{\nabla} \times \vec{B} = \zeta \vec{B}$ where $1/\zeta = \lambda_{B}$ has dimensions of a length characterizing the spatial extension of the solution. The constant-$\zeta$ solutions represent the lowest state of hypermagnetic energy which a closed system may attain also in the case where anomalous currents are present, provided the ambient plasma is perfectly conducting \cite{five}. The configurations minimizing the energy density with the constraint that the magnetic helicity be conserved coincide then with the ones obtainable in ideal magnetohydrodynamics (i.e. 
without anomalous currents). This observation explains why it is possible to derive hypermagnetic knot solutions in a hot plasma from their magnetic counterpart \cite{five} (see, in this respect, \cite{three,seven,sevena}).  

The gyrotropic configurations minimizing Eq. (\ref{secd}) exclude the backreaction on the flow. Indeed, because of the smallness of the hyperelectric fields 
the Ohmic current is given by $\vec{J} = \vec{\nabla}\times \vec{B}/(4\pi)$;  at the same time, $\vec{\nabla} \times \vec{B} = \zeta \vec{B}$.
Putting together the two previous observations the second term at the right hand side of Eq. (\ref{secb}) cancels exactly. Consequently HK have the unique property of allowing resistive decay of the field without introducing stresses on the evolution of the bulk velocity of the plasma. Thus, the maximal frequency of the spectrum is only be determined by the conductivity and it is given by\cite{three,five}:
\begin{equation}
\nu_{\sigma} = 58.28 \biggl(\frac{T_{ew}}{10^{2}\, \mathrm{GeV}} \biggr)^{1/2} \biggl(\frac{g^{\prime}}{0.3}\biggr)^{-1}\biggl(\frac{N_{eff}}{106.75}\biggr)^{1/4}\, \mathrm{kHz},
\label{condfreq}
\end{equation}
where $g^{\prime}$ is the hypercharge coupling constant. Equation (\ref{condfreq}) is obtained by computing the comoving wavenumber $k_{\sigma}$ corresponding to the hypermagnetic diffusivity. In general terms we have that $k_{\sigma}^{-2} = \int^{\tau} d\tau^{\prime}/(4\pi\sigma)$ however when the 
comoving conductivity is constant  the previous expression simplifies as $k_{\sigma} \simeq \sqrt{\sigma {\mathcal H}_{ew} }$. 

The role played by HK in the problem of baryogenesis is mainly related to the possible production of hypermagnetic gyrotropy \cite{three,four}
and it might be useful to elucidate a bit the guiding logic of this proposal.
Since the seminal work of Sakharov \cite{sak} various attempts have been made to account for the baryon asymmetry 
of the Universe (BAU in what follows). The standard lore of baryogenesis (see the first article of Ref. \cite{CKN}) stipulates that during a strongly 
first-order electroweak phase transition the expanding bubbles are nucleated while the baryon number 
is violated by sphaleron processes. Given the current value of the Higgs mass, to produce a sufficiently strong (first-order) phase transition and to get enough $CP$ violation at the bubble wall, the standard electroweak theory must be appropriately extended. The second complementary lore for the generation of the BAU is leptogenesis (see the second article of Ref. \cite{CKN}) which can be conventionally realized thanks to heavy Majorana neutrinos decaying out of equilibrium and producing an excess of lepton 
number ($L$ in what follows). The excess in $L$ can lead to the observed baryon number thanks to sphaleron interactions violating $(B+ L)$. 
An admittedly less conventional perspective stipulates that the BAU could be the result of the decay of maximally helical configurations of the hypercharge field  \cite{three}. Indeed, while the $SU_{L}(2)$ anomaly is typically responsible for $B$ and $L$ non-conservation via instantons and sphalerons, the $U_{Y}(1)$ anomaly might lead to the transformation 
of the infra-red modes of the hypercharge field into fermions \cite{three,fivea,sevena}. As suggested in Ref. \cite{three} the production of the BAU demands, in this context, the dynamical generation of the hypermagnetic gyrotropy ${\mathcal K}^{(B)}(\vec{x},\tau) = \vec{B}\cdot\vec{\nabla}\times \vec{B}$. The magnetic and kinetic gyrotropies 
play a crucial role in the mean-field dynamo \cite{vain} and the same notion occurs, with the due differences, in the present context.

The production of a non-vanishing ${\mathcal K}^{(B)}$ can be obtained, for instance, by studying the effective action describing the interaction of a dynamical pseudoscalar with hypercharge fields as proposed in \cite{three} (see also \cite{ax3})
\begin{equation}
S= \int d^{4} x \sqrt{-g} \biggl[ \frac{1}{2}g^{\alpha\beta}
\partial_{\alpha}\psi \partial_{\beta}\psi  - V(\psi) -
\frac{1}{4}Y_{\alpha\beta}Y^{\alpha\beta} + 
c\frac{\psi}{4 M}
Y_{\alpha\beta}\widetilde{Y}^{\alpha\beta}\biggr],
\label{action1}
\end{equation}
where $Y_{\alpha\beta}$ and $\widetilde{Y}^{\alpha\beta}$ are, respectively, the hypercharge field strength and its dual.
Note that $c = c_{\psi Y} \alpha'/(2 \pi)$ where $\alpha'=
g'^2/4\pi$ and $c_{\psi Y}$ is a numerical factor of order $1$; in case $\psi$ would coincide with a conventional  axion, $c_{\psi Y}$ 
could be computed from the Peccei-Quinn charges of all fermions present in the model.
Even more recently this action has been generalized as:
\begin{equation}
S = \int d^4 x \, \sqrt{- g} \biggl[ {\mathcal M}_{\sigma}^{\rho}(\varphi,\psi) 
Y_{\rho\alpha}\, Y^{\sigma\alpha} - {\mathcal N}_{\sigma}^{\rho}(\varphi,\psi) \widetilde{Y}_{\rho\alpha}\, \widetilde{Y}^{\sigma\alpha}
+ {\mathcal Q} _{\sigma}^{\rho}(\varphi,\psi) Y_{\rho\alpha}\, \widetilde{Y}^{\sigma\alpha}\biggr],
\label{action2}
\end{equation}
The symmetric tensors ${\mathcal M}_{\sigma}^{\rho}(\varphi,\psi)$,  ${\mathcal N}_{\sigma}^{\rho}(\varphi,\psi)$  and 
${\mathcal Q} _{\sigma}^{\rho}(\varphi,\psi)$ contain the couplings of the hypercharge either to the inflaton field 
itself (be it for instance $\varphi$) or to some other spectator field (be it for instance $\psi$).  The typical derivative coupling arising in the relativistic theory of Casimir-Polder and Van der Waals interactions \cite{such} is implicitly contained in Eq. (\ref{action2}) when  ${\mathcal Q} _{\sigma}^{\rho}(\varphi,\psi)=0$: in this case Eq. (\ref{action2}) offers a viable framework for inflationary magnetogenesis  \cite{SUSC1} characterized by unequal electric and magnetic susceptibilities. To be even more specific, following the notations of the fourth paper of Ref. \cite{four} the three symmetric tensors appearing in Eq. (\ref{action2}) can be parametrized as follows:
\begin{eqnarray}
 {\mathcal M}^{\lambda}_{\,\, \rho} &=& - \frac{\lambda}{4} \delta^{\lambda}_{\,\rho} - \frac{\lambda_{E}(\varphi, \psi)}{4}\,  \, u^{\lambda} \, u_{\rho}, \qquad  {\mathcal N}^{\lambda}_{\,\, \rho}  = - \frac{\lambda_{B}(\varphi, \psi)}{4}\, \overline{u}^{\lambda} \, \overline{u}_{\rho},
\nonumber\\
{\mathcal Q}^{\lambda}_{\,\, \rho} &=&  \frac{1}{4}\,  [\lambda_{1}(\varphi,\psi) \delta^{\lambda}_{\rho} + \lambda_{2}(\varphi,\psi) \, \overline{u}^{\lambda} \, \overline{u}_{\rho}],
\label{POS}
\end{eqnarray}
where $u_{\rho} = \partial_{\rho}\varphi/\sqrt{g^{\alpha\beta} \partial_{\alpha} \varphi \partial_{\beta} \varphi}$ and 
$\overline{u}_{\rho} = \partial_{\rho}\psi/\sqrt{g^{\alpha\beta} \partial_{\alpha} \psi \partial_{\beta} \psi}$ are the normalized 
gradients of the scalar fields. When $u_{\mu}\to 0$ and $\overline{u}_{\mu}\to 0$ Eq. (\ref{action2}) reduces to Eq. (\ref{action1}).

In spite of the specific mechanism for the production of ${\mathcal K}^{(B)}$ the contribution of the hypermagnetic gyrotropy determines the comoving baryon to entropy ratio $\eta_{B} = n_{B}/\varsigma$ \cite{three}:
\begin{equation}
\eta_{B}(\vec{x},\tau) = \frac{ 3 \alpha' n_{f}}{8\pi \, H} \biggl(\frac{T}{\sigma}\biggl) \frac{{\mathcal K}^{(B)}(\vec{x}, \tau)}{a^4 \rho_{crit}},
\label{BAU}
\end{equation}
where $\varsigma= 2 \pi^2 T^3 N_{eff}/45$ is the entropy density of the plasma
  and  $n_{f}$ is the number of fermionic generations; $N_{eff} = 106.75$ in the standard electroweak theory. Equation (\ref{BAU}) holds when the rate of the slowest reactions in the plasma (associated with the right-electrons) is larger than the dilution rate caused by the hypermagnetic field itself \cite{three}. 

The basic ideas discussed in the present section have been developed by different authors within slightly different approaches.
An incomplete list of references can be found in Ref. \cite{four}. The common aspect of the ideas presented in \cite{four} 
is that gyrotropic configurations of the hypermagnetic field can seed, in some way, the baryon or lepton 
asymmetry. This idea of hypermagnetic baryogenesis or leptogenesis is exactly the one suggested in \cite{three} even if 
the peculiar production mechanism of the gauge fields can be different.  

\newpage

\renewcommand{\theequation}{3.\arabic{equation}}
\setcounter{equation}{0}
\section{Gravitational waves from hypermagnetic knots} 
\label{sec3}

The intermediate frequency range defined by $\nu_{ew}$ and $\nu_{\sigma}$ encompasses the operating window of space-borne interferometers (such as the (e)Lisa or the Bbo/Decigo projects \cite{space}) and the frequencies where terrestrial wide-band interferometers (such as the Ligo/Virgo experiment \cite{ligo}) are operating today (i.e. between few Hz and $10$ kHz).  Space-borne interferometers [such as (e)Lisa (Laser Interferometer Space Antenna), Bbo (Big Bang Observer), and Decigo (Deci-hertz Interferometer Gravitational Wave Observatory)] might operate between few mHz and the Hz hopefully after 2032. Even in the most optimistic case the sensitivities of these instruments will not be immediately 
relevant for the stochastic backgrounds of cosmological origin. In particular the spectral energy density of the inflationary 
signal at intermediate frequency seems to be out of reach. The same might not be true for the signal coming from 
maximally gyrotropic configurations of the hypermagnetic field. To determine the template family for the emission of gravitational radiation we shall first analyze the case of a single knot configuration and then move to the case of stochastic collection of knots. 

In a conformally flat geometry of Friedmann-Robertson-Walker type the background metric is given by $\overline{g}_{\mu\nu} = a^2(\tau)  \eta_{\mu\nu}$ where $a$ is the scale 
factor, $\tau$ the conformal time coordinate and $\eta_{\mu\nu}$ is the Minkowski metric with signature mostly minus, i.e. $(+, \, -,\, -,\,-)$. 
In this background the amplitude of the tensor fluctuations of the geometry is defined as 
\begin{equation}
\delta_{t} g_{ij} = - a^2 h_{ij}, \qquad \partial_{i}h^{ij} = h_{i}^{i} =0,
\label{TT}
\end{equation}
i.e. $h_{ij}$ is, respectively, divergenceless and traceless. Moreover, given a triplet of mutually orthogonal unit vectors 
(e.g. $\hat{m}$, $\hat{n}$ and $\hat{q}$) the two tensor polarzations are defined as:
\begin{equation}
e_{ij}^{\oplus}(\hat{q})= (\hat{m}_{i} \hat{m}_{j} - \hat{n}_{i} \hat{n}_{j}), \qquad e_{ij}^{\otimes}(\hat{q}) = (\hat{m}_{i} \hat{n}_{j} + \hat{n}_{i}\hat{m}_{j}).
\label{pol1}
\end{equation}
For the present ends the sum over the polarizations can be written as
\begin{equation} 
\sum_{\lambda} e^{(\lambda)}_{ij} \,e^{(\lambda)}_{mn} = p_{i\,m}p_{j\,n} + p_{i\,n} p_{j\,m} - p_{i\,j} p_{m\,n}, 
\label{pol2}
\end{equation}
where $p_{ij} = (\delta_{ij} -\hat{q}_{i} \hat{q}_{j})$ is the transverse projector. 
By perturbing the Einstein-Hilbert action coupled to the sources to second order in the amplitude of the tensor modes of the geometry in a conformally flat FRW background  we obtain \cite{ford}:
\begin{equation}
S_{gw} = \int d^{3} x \, d\tau \sqrt{-\overline{g}} \biggl[  \frac{1}{8\ell_{\mathrm{P}}^2} \overline{g}^{\alpha\beta}
\partial_{\alpha} h_{ij} \partial_{\beta} h^{ij} - \frac{1}{2}\Pi_{ij} h^{ij} \biggr],\qquad \ell_{P} =\sqrt{8 \pi G},
\label{fourth}
\end{equation}
where $\Pi_{ij}$ is the anisotropic stress produced by the HK; note that $\overline{g}$ is the determinant of the background metric $\overline{g}_{\mu\nu}$.  After introducing the tensor normal mode $\mu_{ij} = a h_{ij}$ \cite{one}, 
the evolution equations derived from Eq. (\ref{fourth}) can be expressed as:
\begin{equation}
\mu_{ij}^{\prime\prime} - \nabla^2 \mu_{ij} - \frac{a^{\prime\prime}}{a} \mu_{ij} = - 2 \ell_{P}^2 a^3(\tau) \Pi_{ij}, \qquad \mu_{ij} = a h_{ij}.
\label{gw2}
\end{equation}
As previously remarked, the backreaction on the flow vanishes because the Ohmic current and the hypermagnetic field are parallel. Thus the HK the solution can be factorized as 
\begin{equation}
\vec{B}(\vec{x},\tau) = \vec{b}(\vec{x}) f(\zeta,\tau), \qquad f(\zeta,\tau)= \exp{[ - \zeta^2 \tau/(4 \pi\sigma) ]}
\label{gw2a}
\end{equation}
where $\vec{b}(\vec{x})$ minimizes Eq. (\ref{secd}) and $f(\zeta,\tau)$
accounts for the contribution of the Ohmic dissipation in the case of a single configuration with constant $\zeta$. According to the well known Chandrasekhar-Kendall representation \cite{six},  $\vec{b}(\vec{x})$ can always be expressed as:
\begin{equation}
\vec{b}(\vec{x}) = \lambda_{B} \vec{\nabla}\times [ \vec{\nabla} \times ( \hat{u} \Psi)] + \vec{\nabla} \times  ( \hat{u} \Psi),
\label{third}
\end{equation}
where $\hat{u}$ is a unit vector denoting the direction of the knot and $\Psi$ obeys $\nabla^2 \Psi+ \zeta^2 \Psi=0$. 
If $\zeta$ is constant, the general representation of the hypermagnetic knot configuration can be achieved through the Chandrasekhar-Kendall representation \cite{six}. Equation (\ref{third}) leads to a gyrotropy that does not decrease at large distance scales but more realistic configurations where the gyrotropy does decrease at large distance scales can be found \cite{three,sixa}.

 The preferred direction of the knot introduces a difference in the evolution of the two polarizations of the gravitational wave. Indeed, 
 the anisotropic stress associated with the 
HK is given by $\Pi_{ij}= (B_{i} B_{j} - B^2 \delta_{ij}/3)/(4 \pi a^4(\tau))$ and it can always be projected along the two tensor polarizations as $\Pi_{ij} =  (\Pi_{\oplus}e_{ij}^{\oplus}+  \Pi_{\otimes} e_{ij}^{\otimes})/2$ where $\Pi_{\oplus}$ and $\Pi_{\otimes}$ are, respectively,
\begin{equation}
\Pi_{\oplus}(\vec{x},\tau) = \frac{1}{4\pi\,a^4(\tau)}\biggl[(\vec{B}\cdot\hat{m})^2 - (\vec{B}\cdot\hat{n})^2\biggr], \qquad \Pi_{\otimes}(\vec{x},\tau) = \frac{1}{2\pi\,a^4(\tau)} (\vec{B}\cdot\hat{m}) (\vec{B}\cdot\hat{n}).
\label{gw5}
\end{equation}
Since the direction of the knot does not need to coincide with the direction of propagation of the gravitational wave,  Eq. (\ref{gw5}) implies that the emission is polarized. When the electroweak epoch represents a portion of the radiation-dominated evolution (as assumed throughout) the scale factor evolves linearly in conformal time; thus $a^{\prime\prime}=0$ in Eq. (\ref{gw2}) and the corresponding solution is:
\begin{eqnarray}
h_{ij}(\vec{x},\tau) &=& - \frac{ 2 \ell_{P}^2}{a(\tau)} \int d^{3} x^{\prime} \, \int_{\tau_{ew}}^{\tau} d\xi \,{\mathcal G}(\vec{x}, \vec{x}^{\,\prime}; \tau, \xi) \, a^3(\xi) \, \Pi_{ij}(\vec{x}^{\prime}, \xi),
\nonumber\\
{\mathcal G}(\vec{x}, \vec{x}^{\,\prime}; \tau, \xi) &=& \frac{1}{(2\pi)^3} \int\frac{ d^{3} k}{k} \,e^{- i \vec{k}\cdot(\vec{x} - \vec{x}^{\,\prime})} \, \sin{[ k (\xi - \tau)]}.
\label{gw6}
\end{eqnarray}
Thanks to Eq. (\ref{third}) also the whole anisotropic stress of the HK can be  factorized 
as $\Pi_{ij}(\vec{x},\tau)= \overline{\Pi}_{ij}(\vec{x}) f(\sqrt{2}\zeta,\tau)/a^4(\tau)$.
Using this observation and recalling that $ a {\mathcal H} =  a_{ew} {\mathcal H}_{ew}$ is constant during radiation we have:
\begin{eqnarray}
h_{ij}(\vec{k}, \tau) &=& - \frac{2 \ell_{P}^2 \tau_{ew}}{a(\tau)} \int d^{3} k \,\overline{\Pi}_{ij}(\vec{k})\, {\mathcal F}(k\tau) 
e^{-i \vec{k}\cdot \vec{x}},
\nonumber\\
{\mathcal F}(k\tau) &=& \int_{1}^{k\tau} d y \frac{ \sin{(y - k\tau)}}{y} \biggl[ 1 -  {\mathcal O}\biggl(\frac{\zeta^2 y}{\sigma k}\biggr)\biggr], \qquad y = k \tau_{ew}.
\label{gw7}
\end{eqnarray}
Since we must integrate over all the modes inside the particle horizon 
at $\tau_{ew}$  the lowest extremum of integration in Eq. (\ref{gw7}) coincides with $1$. 
To simplify the integrand even further we are just considering  those scales which are not effected by the hypermagnetic 
diffusivity. Thus $f(\sqrt{2}\zeta,\tau)$ can be expanded by neglecting all higher-order corrections and by positing, as it is in practice, that the signal is absent for scales smaller than the diffusivity scale.  The approximate expression of ${\mathcal F}(k\tau)$ appearing in Eq. (\ref{gw7}) can then be written as:
\begin{equation}
{\mathcal F}(k\tau) = \sin{k \tau} [ \alpha - \mathrm{Ci}(k\tau)] + \cos{k\tau} [ \mathrm{Si}(k\tau) -\beta],
\label{gw8}
\end{equation}
where $\alpha = \mathrm{Ci}(1) =0.33$ and $\beta =  \mathrm{Si}(1) =0.94$. Note that, in Eq. (\ref{gw8}), the usual trigonometric intergrals 
are defined as 
\begin{equation}
\mathrm{Ci}(z)= - \int_{z}^{\infty} dt \frac{\cos{t}}{t}, \qquad \mathrm{Si}(z) =  \int_{z}^{\infty} dt \frac{\sin{t}}{t}.
\label{gw8a}
\end{equation}
In the limit $k\tau \gg 1$ we have that ${\mathcal F}(k\tau) \to (\pi/2 - \beta) \cos{k\tau} + \alpha \sin{k\tau} $; this expression
also implies $|{\mathcal F}(k\tau)| \leq \epsilon$ where $\epsilon = 0.7$. 

Having determined the emitted amplitude in the case of a single HK, we can now compute the energy density and, for this 
purpose, the energy-momentum tensor of the gravitational waves in a FRW background can be formally
derived by varying the free part of the action (\ref{fourth}) with respect to $\overline{g}^{\alpha\beta}$ and the result is:
\begin{equation}
{\mathcal T}_{\mu}^{\nu} = \frac{1}{4\ell_{\mathrm{P}}^2}\biggl[ \partial_{\mu} h_{ij} \partial^{\nu} h^{ij} - \frac{1}{2} \,\delta_{\mu}^{\nu} \,
\overline{g}^{\alpha\beta} \,\partial_{\alpha} h_{ij} \partial_{\beta} h^{ij} \biggr].
\label{gw9}
\end{equation}
This expression due to Ford and Parker \cite{ford} relies on the physical 
observation that the two polarizations of the gravitational waves in a conformally flat background behave as a pair of minimally coupled scalar fields.  
However, since the energy and momentum of the gravitational field itself cannot be localized there is no unique expression for ${\mathcal T}_{\mu\nu}$, we must instead deal with pseudotensors whose definitions can be mathematically slighty different but are physically equivalent \cite{sixb}. Indeed it has been argued in the past that different computational schemes  lead exactly to the same spectral energy density for wavelengths shorter than the Hubble radius at each corresponding epoch. Note that the energy-momentum pseudo-tensor defined from the second-order variation of the Einstein tensor and the  expression of Eq. (\ref{gw9}) are formally slightly different and their peruse in the context of stochastic backgrounds of relic gravitons has been carefully discussed in \cite{sixb} (see, in particular, the last paper). 

With these necessary precisions the energy density defined from Eq. (\ref{gw9}) is given by:
\begin{equation}
\rho_{gw} = \frac{1}{8 \ell_{P}^2 a^2 } \biggl[ \partial_{\tau} h_{ij} \partial_{\tau} h_{ij} + \partial_{k} h_{ij} \partial_{k}h_{ij} \biggr].
\label{gw10}
\end{equation}
Since the first term at the right hand side of  Eq. (\ref{gw10}) is always subleading when the relevant modes are shorter 
than the particle horizon the energy density in critical units becomes: 
\begin{equation}
\frac{\rho_{gw}}{\rho_{cr}} = \frac{{\mathcal N}}{(1+z_{eq})(1+ z_{\Lambda})^3} \frac{b^4}{(3 H_{ew}^2 \overline{M}_{P}^2)^2}, \qquad {\mathcal N} = \frac{\epsilon^2}{16 \pi^2},
\label{gw11}
\end{equation}
where $\overline{M}_{P}= \ell_{P}^{-1}$ and $\overline{\Pi}_{ij}(\vec{x}) \overline{\Pi}^{ij}(\vec{x}) = b^4/(24\pi^2)$.
Note that $1+ z_{eq} = 3228.9 (h_{0}^2 \Omega_{M0}/0.134)$ is the redshift to equality while $1 + z_{\Lambda} = 0.703 (\Omega_{M0}/0.258)^{1/3} (\Omega_{\Lambda}/0.742)^{-1/3}$ is the redshift of $\Lambda$ dominance (recall that in the concordance model the dark energy component is simply parametrized in terms of a cosmological constant). Note that, after equality, the energy density of the gravitational waves still redshifts like radiation while the background is dominated by matter; after the moment of $\Lambda$-dominance the background energy density is instead constant. The overall result of these effects leads to the final form of Eq. (\ref{gw11}).  

Equation (\ref{gw11}) gives the energy density emitted by a single HK in real space. A stochastic collection of HK is characterised, in Fourier space,  by the two-point function:
\begin{equation}
\langle B_{i}(\vec{k},\tau) \, B_{j}(\vec{p},\tau^{\prime}) \rangle = \frac{2\pi^2}{k^3} \epsilon_{ijk} \hat{k}^{k} P_{hk}(k,\tau,\tau^{\prime}) \, \delta^{(3)}(\vec{k} + \vec{p}),
\label{gw12}
\end{equation}
where the power spectrum $P_{hk}(k,\tau)$ has the dimensions of an energy density. As in the case of a single knot the
resistive decay of the field does not introduces stresses on the evolution of the bulk velocity. Equations (\ref{gw7}) and (\ref{gw10}) 
can still be used together with Eq. (\ref{gw12}) for a direct computation of the spectral energy density in critical units\footnote{In the present paper $\ln{}$ denotes the natural logarithm while $\log{}$ denotes the common logarithm.}:  
\begin{eqnarray}
\Omega_{gw}(k, \tau) &=& \frac{1}{\rho_{\mathrm{crit}}} \frac{d \rho_{gw}}{d \ln{k}} = \frac{\ell_{P}^4}{6 a ^4 H^2} 
\int_{\tau_{ew}}^{\tau} d\xi \, \int_{\tau_{ew}}^{\tau} d\xi^{\prime} a^3(\xi) a^3(\xi^{\prime}) 
\nonumber\\
&\times& \sin{[k (\xi - \tau)]} \sin{[k (\xi^{\prime} - \tau)]} P_{\Pi}(k,\xi, \xi^{\prime})
\biggl[ 1 + {\mathcal O}\biggl(\frac{{\mathcal H}^2}{k^2}\biggr)\biggr],
\label{gw13}
\end{eqnarray}
where the subleading piece (second term in the squared bracket) comes from the time derivative of $h_{ij}$ appearing in Eq. (\ref{gw10}) which is negligible inside the Hubble radius. The power spectrum of the anisotropic stress appearing in Eq. (\ref{gw13}) is quadratic in the power spectra of the HK and its explicit form is given by:
\begin{eqnarray}
P_{\Pi}(q, \xi, \xi^{\prime}) = \frac{5 q^3 }{96 \pi^3 a^{4}(\xi) a^{4}(\xi^{\prime})} \int d^{3} k \frac{(1 - \hat{k}\cdot\hat{q})}{k^3 |\vec{q} - \vec{k}|^3} 
P_{hk}(k, \xi,\xi^{\prime}) P_{hk}(|\vec{q} - \vec{k}|, \xi,\xi^{\prime}).
\label{gw14}
\end{eqnarray}
The spectral energy density of Eq. (\ref{gw13}) in critical units is dimensionless but scales with the square of the amplitude of the knot power spectrum. Equation 
(\ref{gw14}) has therefore the same physical content of Eq. (\ref{gw11}) where $\rho_{gw}/\rho_{crit}$ scales with the square of the energy density 
of the HK.  Notice, incidentally, that to derive Eq. (\ref{gw14}) we need to evaluate the correlator of the anisotropic stresses. In particular 
after some algebra it is relatively straightforward to arrive at the following expression:
\begin{eqnarray}
&& \langle \Pi_{ij}(\vec{q}, \xi)\, \Pi_{ij}(\vec{q}^{\,\prime}, \xi^{\prime}) \rangle = \frac{1}{128 \pi^{5} a^{4}(\xi) a^{4}(\xi^{\,\prime})} \int d^{3} k  \,\int d^{3} p \int d^{3} k^{\prime}  \,\int d^{3} p^{\prime} 
\nonumber\\
&& \times \delta^{(3)}(\vec{q} - \vec{k} - \vec{p}) 
\delta^{(3)}(\vec{q}^{\,\prime} - \vec{k}^{\,\prime} - \vec{p}^{\,\prime}) \biggl[ \langle B_{i}(\vec{k}, \xi) B_{j}(\vec{p}, \xi) B_{i}(\vec{k}^{\,\prime}, \xi^{\prime}) B_{j}(\vec{p}^{\,\prime}, \xi^{\prime}) \rangle
\nonumber\\
&& - \frac{1}{3}\langle B_{\ell}(\vec{k}, \xi) B_{\ell}(\vec{p}, \xi) B_{m}(\vec{k}^{\,\prime}, \xi^{ \prime}) B_{m}(\vec{p}^{\,\prime}, \xi^{\prime}) \rangle
\biggr].
\label{gw14a}
\end{eqnarray}
The different correlation functions appearing in Eq. (\ref{gw14a}) can be computed explicitly in terms of the two-point 
functions of Eq. (\ref{gw12}). After some algebra the result of Eq. (\ref{gw14}) can be swiftly obtained by recalling that, according to the 
present notations, the power spectrum of the anisotropic stresses is defined as $\langle \Pi_{ij}(\vec{q}, \xi)\, \Pi_{ij}(\vec{q}^{\,\,\prime}, \xi^{\prime}) \rangle= (2 \pi^2/k^3 )\delta^{(3)}(\vec{q}+ \vec{q}^{\,\,\prime}) \, P_{\Pi}(q,\,\xi,\,\xi^{\prime})$. 

As in the case of Eq. (\ref{third}) the evolution can be factorized and also the power spectrum can be written as 
$P_{hk}(k,\tau) = A_{hk} (k/k_{ew})^{\beta} e^{- k^2(\tau+\tau^{\prime})/(4 \pi \sigma)}$. The power-law is the simplest parametrization for the power spectrum and it appears in the context of different models (see e.g. \cite{three,four,five,sevena}).  Thus from Eq. (\ref{gw13}) the spectral energy density 
can be expressed as:  
 \begin{eqnarray}
 \Omega_{gw}(q, \tau_0) &=& \frac{5 \pi \ell_{P}^4 A_{hk}^2 }{288 (1+z_{eq})( 1+ z_{\Lambda})^3 H_{ew}^4} {\mathcal F}^2 (q\tau_{ew}) \biggl(\frac{q}{q_{ew}}\biggr)^{2\beta} e^{ -2 q^2/q_{\sigma}^2} {\mathcal Q}(q, q_{\sigma}, q_{ew}),
 \nonumber\\
 {\mathcal Q}(q, q_{\sigma}, q_{ew}) &=& \biggl\{\frac{2}{\beta} \biggl[ 1 - \biggl(\frac{q_{ew}}{q}\biggr)^{\beta}\biggr] + \frac{2(\beta-3)}{3 (\beta+1)}\biggl[ 1 -\biggl(\frac{q_{ew}}{q}\biggr)^{\beta+1}\biggr]
 \nonumber\\
 &-& \frac{2}{3 - 2\beta}\biggl[ \biggl(\frac{q}{q_{\sigma}}\biggr)^{3 - 2\beta} -1\biggr] + \frac{(\beta -3)}{3 ( 2 - \beta)} \biggl[\biggl(\frac{q}{q_{\sigma}}\biggr)^{4 - 2 \beta} -1\biggr]\biggr\}
 \label{gw15}
 \end{eqnarray}
Equation (\ref{gw15}) holds, strictly speaking, for $ 0 <\beta <1.5$; when $\beta \to 0$ and $\beta \to 1.5$ the explicit expression of ${\mathcal Q}$ will contain some extra logarithms.  In the range\footnote{Needless to say that we can easily pass from wavenumbers to frequencies since $2\pi \nu = k$, as already mentioned in section \ref{sec1}.}  of frequencies 
$\nu_{ew}< \nu< \nu_{\sigma}$  we have that $5 \pi \epsilon^2 {\mathcal Q}(\nu, \nu_{\sigma}, \nu_{ew})/288$ varies between $0.01$ and $0.8$.
The same approximate values hold also in the limits $\beta\to 0$ and $\beta\to 1.5$.
As long as the overall amplitude is not divergent of frequencies (as we just showed), 
the specific value of the numerical factors is not strictly essential for the determination of the template: 
the numerical factors can always be reshuffled in an overall amplitude which will be eventually constrained by phenomenology and (even more 
optimistically) by direct measurements. With this strategy in mind, from Eq. (\ref{gw15}) the template for the emission of gravitational waves from a 
background of HK is:
\begin{equation}
\Omega_{gw}(\nu,\tau_0)=  \frac{\Omega_{B}^2}{(1 + z_{eq}) (1 + z_{\Lambda})^3}  \biggl(\frac{\nu}{\nu_{ew}}\biggr)^{\alpha} e^{ - 2 (\nu/\nu_{\sigma})^2},\qquad \nu\geq \nu_{ew},
\label{gw18}
\end{equation}
where $\alpha=2\beta$. All the numerical factors have then been reabsorbed into $\Omega_{B}^2$ which is chiefly determined 
by the dimensionless ratio $A_{hk}^2/(H_{ew}^4 \overline{M}_{P}^2)$. 

The template given by Eq. (\ref{gw18}) can be constrained by using all the specific bounds applicable to the 
stochastic backgrounds of gravitational radiation of cosmological and astrophysical origin. As we shall see already by imposing 
these constraints a relevant portion of the parameter space can be excluded. The remaining regions may 
lead to a signal which could be observed, in principle, both by space-borne detectors and by terrestrial interferometers. 
The results of this analysis are reported in the following section.

\renewcommand{\theequation}{4.\arabic{equation}}
\setcounter{equation}{0}
\section{Phenomenological considerations and sensitivities} 
\label{sec4}

To extract specific informations on $\Omega_{B}$ and $\alpha$, Eq. (\ref{gw18}) 
can be compared with the current bounds applying to the 
stochastic backgrounds of relic gravitons. The large-scale bounds (constraining $r_{\mathrm{T}}$ in the aHz region) 
are immaterial  since, according to Eq. (\ref{gw18}), $\nu_{ew}$ is the minimal frequency of the spectrum. 
For the same reason the pulsar timing limits \cite{pulsar} demand $\Omega(\nu_{pulsar},\tau_{0}) < 1.9\times10^{-8}$ for a typical frequency  
$\nu_{pulsar} \simeq \,10\,\mathrm{nHz}$. Since  $\nu_{\mathrm{pulsar}}$ roughly corresponds to the inverse 
of the observation time during which the pulsars timing has been monitored this potential constraint 
is always satisfied by Eq. (\ref{gw18}) since $\nu_{ew} \ll \nu_{pulsar}$.  
\begin{figure}[!ht]
\centering
\includegraphics[height=7cm]{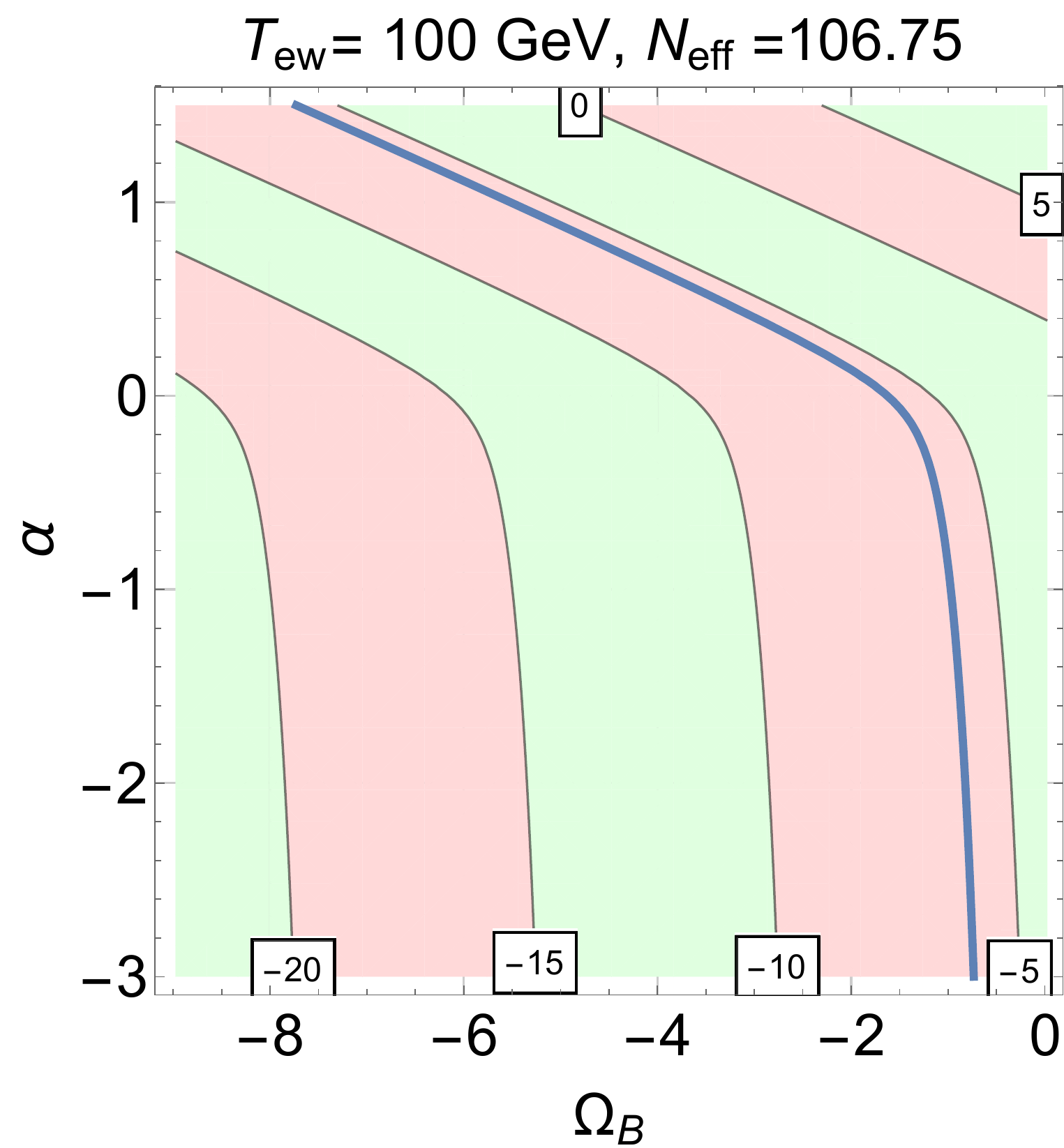}
\includegraphics[height=7cm]{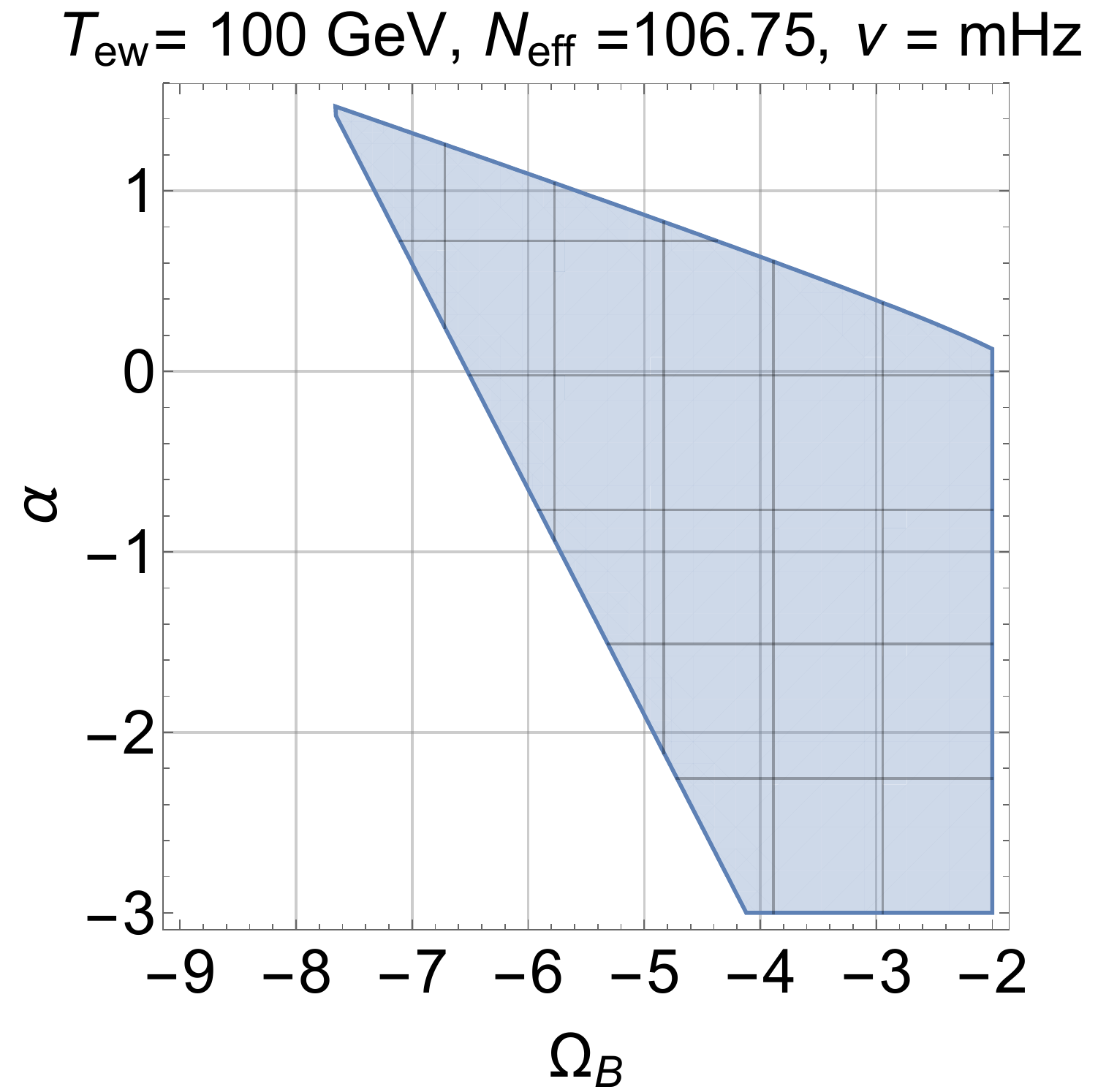}
\caption[a]{In the plot at the left we illustrate the common logarithm of the integral appearing at the left 
hand side of Eq. (\ref{gw19}). The thick line corresponds to the nucleosynthesis bound. In the plot at the right we 
chart the region of the parameter space where $h_{0}^2\Omega_{gw}$ is larger than the inflationary signal 
for a putative frequency of the mHz.}
\label{Figure1}      
\end{figure}
A qualitatively different constraint 
stems from big-bang nucleosynthesis \cite{bbn}. If there are some additional relativistic degrees of
freedom (either bosonic or fermionic) the effect on the expansion rate will be the same as that of having some 
(perhaps a fractional number of) additional neutrino species. Before electron-positron annihilation we have $\rho_X = (7/8)\Delta N_{\nu} \rho_\gamma$
and after electron-positron annihilation we have $\rho_X = (7/8) (4/11)^{4/3} \,\Delta N_{\nu} 
\,\rho_{\gamma} \simeq 0.227\,\Delta N_{\nu} \,\rho_\gamma$. The critical fraction of CMB photons can be directly computed from 
the value of the CMB temperature and it is given by $h_{0}^2 \Omega_\gamma \equiv \rho_\gamma/\rho_{\mathrm{crit}} = 
2.47\times10^{-5}$. If the extra energy density component has stayed radiation-like until today, its ratio to the critical density is given by:
\begin{equation}
 h_{0}^2\frac{\rho_X}{\rho_{\mathrm{c}}} = 5.61\times10^{-6}\Delta N_{\nu} 
\biggl(\frac{h_{0}^2 \Omega_{\gamma0}}{2.47 \times 10^{-5}}\biggr), \qquad p_{X}= \frac{\rho_{X}}{3}.
\label{bbn0}
\end{equation}
In case the additional species are  gravitons, then  Eq. (\ref{bbn0}) implies\cite{bbn}: 
\begin{equation}
h_{0}^2  \int_{\nu_{ew}}^{\infty}
  \Omega_{gw}(\nu,\tau_{0}) d\ln{\nu} = 5.61 \times 10^{-6} \Delta N_{\nu} 
  \biggl(\frac{h_{0}^2 \Omega_{\gamma0}}{2.47 \times 10^{-5}}\biggr).
\label{gw19}
\end{equation}
The bounds on $\Delta N_{\nu}$ range from $\Delta N_{\nu} \leq 0.2$ 
to $\Delta N_{\nu} \leq 1$;  the integrated spectral density is thus between $10^{-6}$ and $10^{-5}$. 
In general terms the lower extremum of integration should coincide with the present frequency corresponding to the 
Hubble rate at the time of big-bang nucleosynthesis; this quantity, conventionally denoted $\nu_{bbn}$,
is of the order of $0.01$ nHz for a putative nucleosynthesis temperature ${\mathcal O}(1)$ MeV 
and for an effective number of relativistic degrees of freedom corresponding to $10.75$. 
In the range $\nu_{bbn} \leq \nu <  \nu_{ew}$  the signal emitted from the HK is absent; therefore the lower extremum of integration 
in Eq. (\ref{gw19}) coincides with the lowest frequency of the spectrum. 
\begin{figure}[!ht]
\centering
\includegraphics[height=7cm]{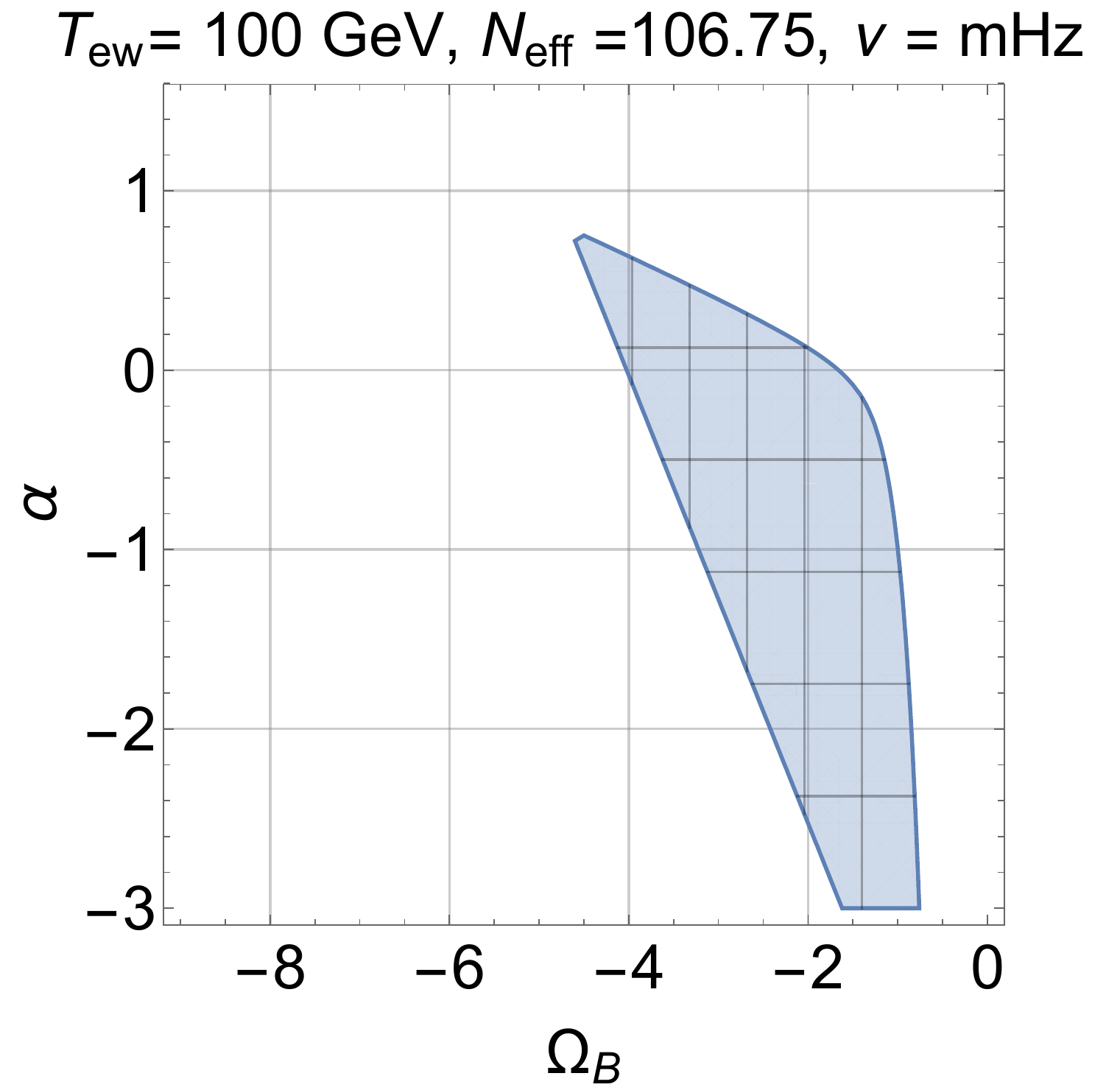}
\includegraphics[height=7cm]{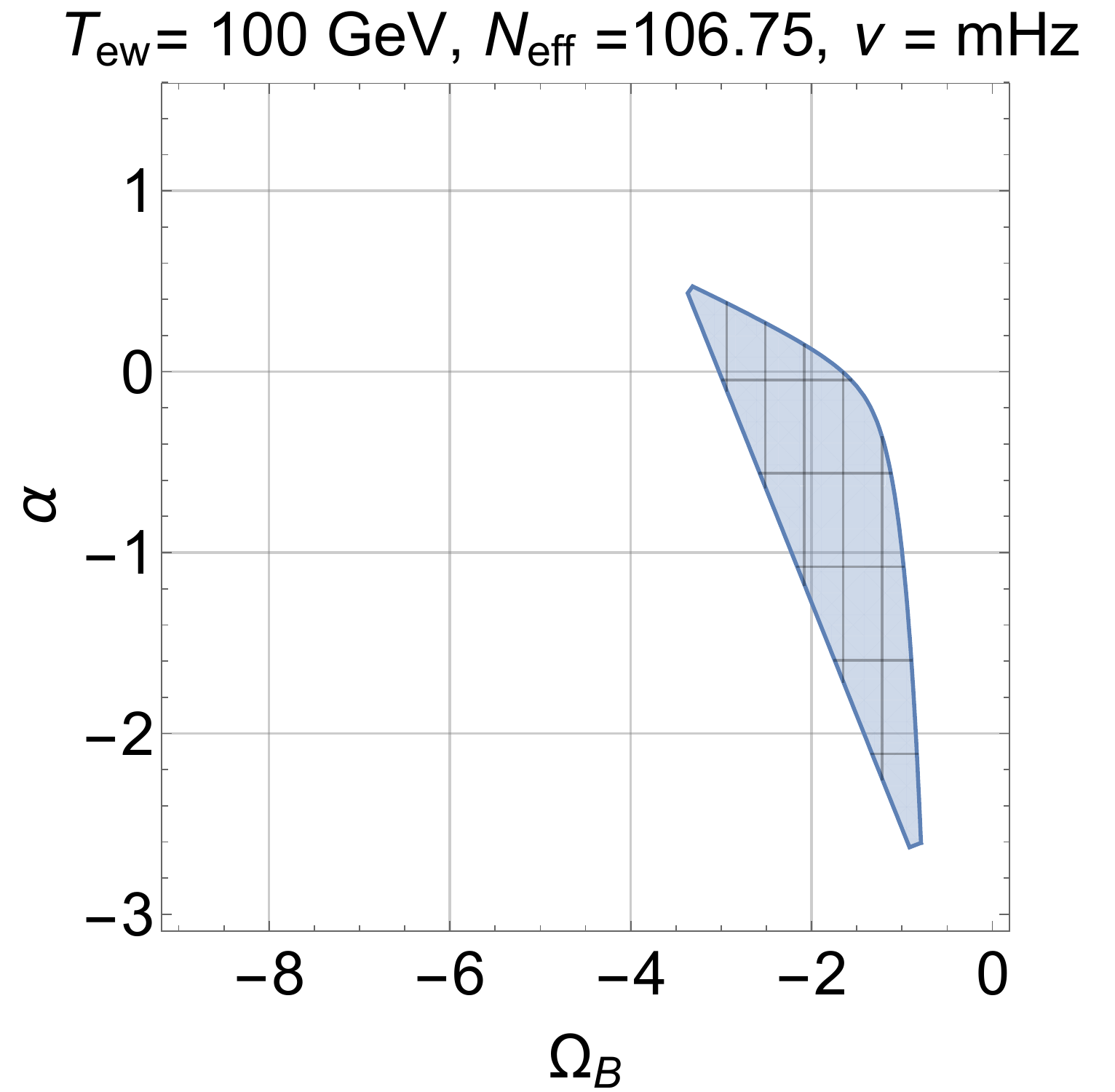}
\includegraphics[height=7cm]{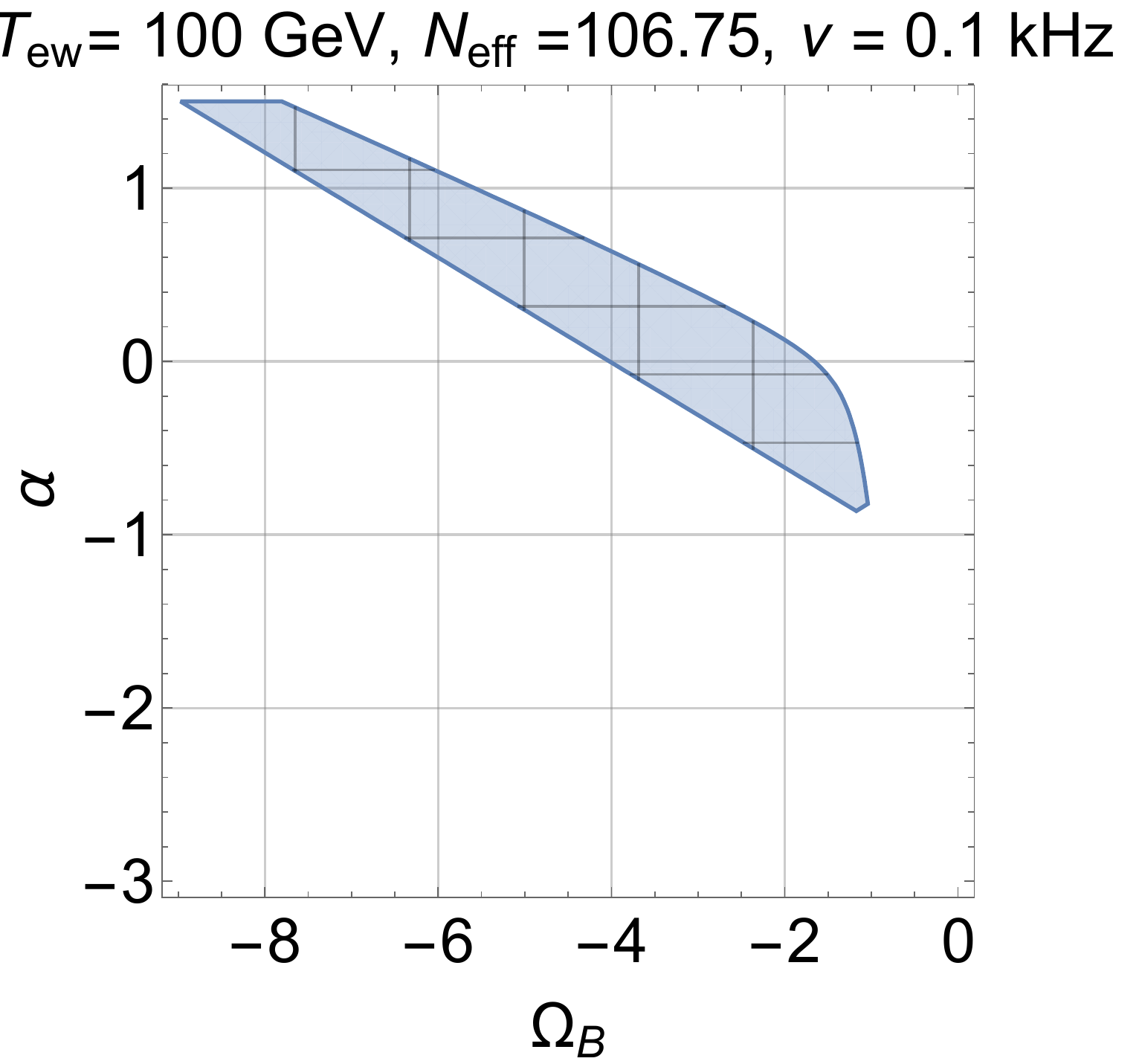}
\includegraphics[height=7cm]{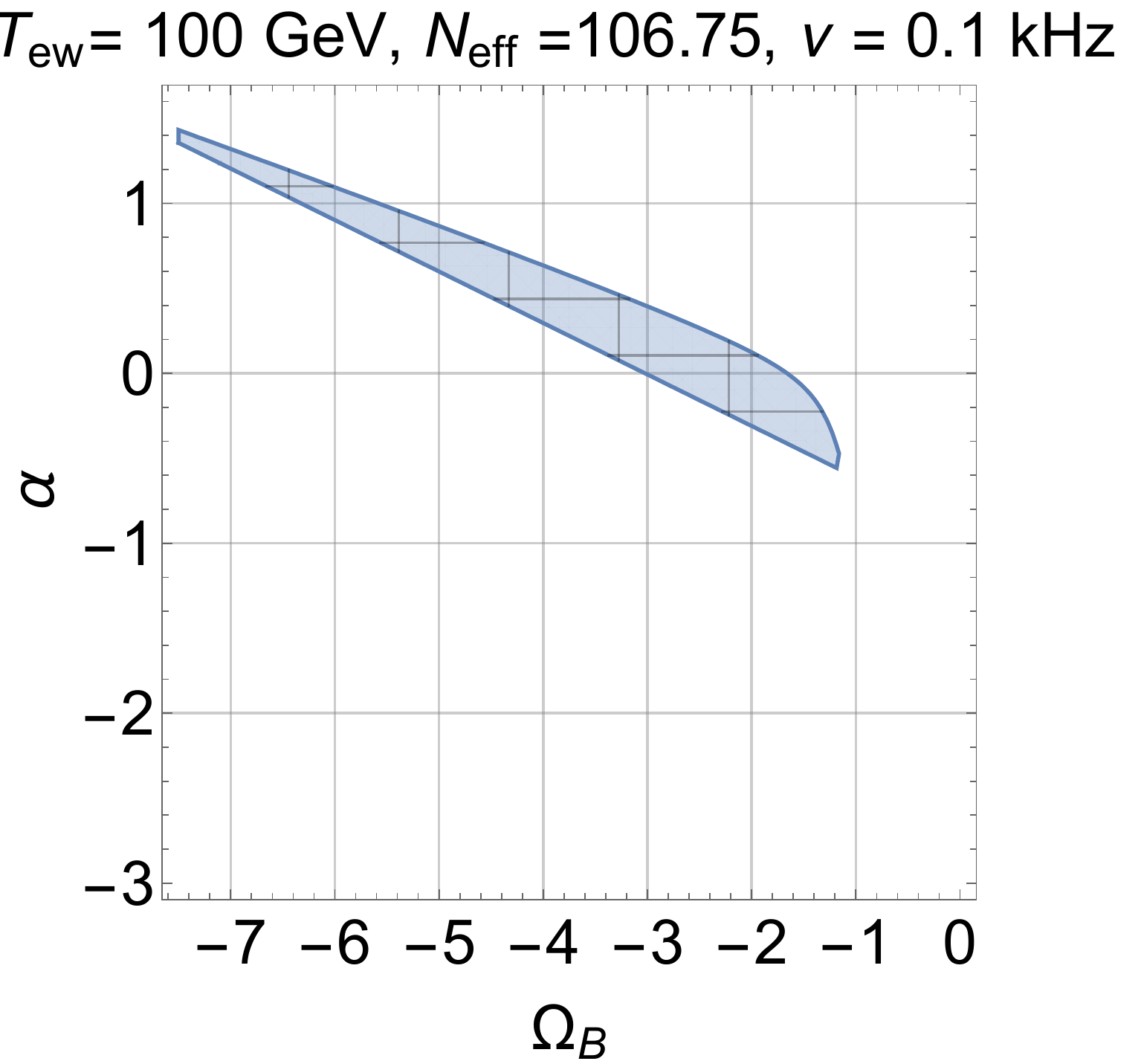}
\caption[a]{The shaded areas in the two plots at the left illustrate the regions accessible to wide-band interferometers 
with putative sensitivity ${\mathcal O}(10^{-12})$. Similarly the two plots at the right the sensitivity 
has been taken ${\mathcal O}(10^{-10})$. The two plots at the top correspond to frequencies 
of the mHz which is the characteristic window of space-borne interferometers \cite{space}. The two plots at the bottom
concern a typical frequency of $0.1$ kHz which is characteristic of terrestrial interferometers \cite{ligo}. }
\label{Figure2}      
\end{figure}
Using Eq. (\ref{gw18}) the integral of Eq. (\ref{gw19}) can be performed explicitly. 
The common logarithm of the left hand side of Eq. (\ref{gw19}) is illustrated in the left plot of Fig. \ref{Figure1}. 
The thick line corresponds to the common logarithm of the right-hand side of Eq. (\ref{gw19}) in the case $\Delta N_{\nu} =0.2$. 
In the plane $(\Omega_{B}, \alpha)$ the region allowed by Eq. (\ref{gw19}) should always be below the thick line. In the plot at the right of Fig. \ref{Figure1} 
the shaded area illustrates the portion of the parameter space where the bound of Eq. (\ref{gw19}) is enforced and, simultaneously, 
$h_{0}^2 \Omega_{gw} > 10^{-16.5}$ for a typical frequency of the order of the mHz characterizing  space-borne interferometers such as (e)Lisa and 
Bbo/Decigo. In the (e)Lisa case we shall assume a nominal sensitivity of $10^{-11}$ (for the spectral energy density)
in a frequency range centered around the mHz. In the Bbo/Decigo case the sensitivity  could even be ${\mathcal O}(10^{-15})$ and for a typical 
frequency range centered around the Hz. All in all, from the right plot of Fig. \ref{Figure2} we can just infer that 
for frequencies around the mHz the signal of HK can be larger than the inflationary signal, 
compatible with the nucleosynthesis constraint of Eq. (\ref{gw19}) and even potentially detectable.

As we move to higher frequencies the portion of the parameter space compatible with the current bounds and 
reachable by the hoped sensitivities of terrestrial and space-borne detectors becomes smaller. This aspect is specifically 
illustrated in the fourfold plot of Fig. \ref{Figure2}. In the two plots at the left we required $h_{0}^2\Omega_{gw}$ to be larger 
than $10^{-12}$ while in the two plots at the right the shaded area illustrates the region where 
$h_{0}^2\Omega_{gw} > 10^{-10}$. In all the plots of Fig. \ref{Figure2} the current constraints discussed above have been 
consistently enforced. The Ligo/Virgo sensitivity to a stochastic background depends on $\alpha$. Unfortunately 
in its current configuration the experiment can only access regions where $h_{0}^2 \Omega_{gw} = {\mathcal O}(10^{-4})$ \cite{ligo}.
These regions have been already excluded thanks to the current phenomenological constraints. Advanced Ligo/Virgo detectors 
might be far more sensitive and probe the shaded area of Fig. \ref{Figure2}.
By looking simultaneously at Figs. \ref{Figure1} and \ref{Figure2} we can conclude that at intermediate 
frequencies $ \mu\mathrm{Hz} \leq \nu \leq \mathrm{kHz}$ the gravitational wave background induced by HK is always 
larger than the inflationary signal even if the allowed region of the parameter space is severely restricted by the 
current phenomenological bounds. Already the terrestrial interferometers (in their most advanced and sensitive configurations) 
will be able to exclude the region of white or blue spectra indices (i.e. $\alpha \geq -1$) and small amplitudes (i.e. $\Omega_{B} < 10^{-4}$).
Conversely the space-borne detectors will preferentially probe the region of large amplitudes 
(i.e. $\Omega_{B} = {\mathcal O}(10^{-3})$) and red spectral indices (i.e. $\alpha < -2$). The regions of red spectral 
indices can be already excluded on a theoretical ground since they would correspond to gyrotropic configurations of the 
hypermagnetic field increasing (rather than decreasing) at large-distance scales. We are therefore 
left with the two tiny filled regions in the two plots at the bottom of Fig. \ref{Figure2}.

\newpage

\renewcommand{\theequation}{5.\arabic{equation}}
\setcounter{equation}{0}
\section{Concluding remarks} 
\label{sec5}
The intermediate frequency range of the spectrum of relic gravitational radiation goes from few $\mu$Hz to $10$ kHz
This intermediate range encompasses the operating windows of space-borne interferometers (hopefully available twenty years from now) 
and of terrestrial detectors (already available but still insensitive to stochastic backgrounds of relic gravitons 
of cosmological origin). This statement can be understood by comparing the quoted 
sensitivities of the Ligo/Virgo experiments with the constraints imposed by the big-bang 
nucleosynthesis bound. Moreover, between few $\mu$Hz and $10$ kHz, the conventional  inflationary models lead to relic gravitons whose spectral energy density can be (at most) of the order of $10^{-17}$.

There are however configurations of the hypermagnetic field carrying both magnetic helicity 
and magnetic gyrotropy, dubbed hypermagnetic knots, producing gravitational radiation 
with a spectrum ranging between $\nu_{ew} = {\mathcal O}(20)\, \mu\mathrm{Hz}$ and 
$\nu_{\sigma} =  {\mathcal O}(50)\, \mathrm{kHz}$. In this paper we showed how to construct a physical template family for the emission of the gravitational radiation produced by the hypermagnetic knots. While between $\nu_{ew}$ and $\nu_{\sigma}$ the inflationary contribution implies 
a spectral energy density $h_{0}^2 \Omega_{gw}^{(inf)} = {\mathcal O}(10^{-17})$,
the signal due to hypermagnetic knots can be as large as  $h_{0}^2 \Omega_{gw}^{(knots)} = {\mathcal O}(10^{-8})$
without conflicting with current bounds applicable to stochastic backgrounds of gravitational radiation.

The lack of observation of gravitational waves between few $\mu$Hz and 
$10$ kHz will potentially exclude the presence of hypermagnetic knots configurations at the electroweak scale. Conversely the 
observation of a signal in the range that encompasses the operating windows of space-borne and terrestrial wide-band 
detectors will not necessarily confirm the nature of the source. Further scrutiny will be needed but the signal of the hypermagnetic knots can be 
disambiguated since the stochastic background of gravitational 
waves produced by the hypermagnetic knots is polarized. Last but not least a gravitational signal coming from maximally 
gyrotropic configurations of the hypercharge may offer an indirect test of the equations of anomalous magnetohydrodynamics whose spectrum includes hypermagnetic knots and Chern-Simons waves as low-frequency excitations.


\newpage

\begin{thebibliography}{99}

\bibitem{one}  L.~P.~Grishchuk,   Sov.\ Phys.\ JETP {\bf 40}, 409 (1975)   [Zh.\ Eksp.\ Teor.\ Fiz.\  {\bf 67}, 825 (1974)]; 
 Annals N.\ Y.\ Acad.\ Sci.\  {\bf 302}, 439 (1977); A.~A.~Starobinsky, JETP Lett.\  {\bf 30}, 682 (1979); V. A. Rubakov, M. V. Sazhin and 
 A. V. Veryaskin, Phys. Lett. B {\bf 115}, 189 (1982); B. Allen, Phys. rev. D {\bf 37}, 2078 (1988);
 V. Sahni, Phys. Rev. D {\bf 42}, 453 (1990);   L.~P.~Grishchuk and M.~Solokhin, Phys.\ Rev.\ D {\bf 43}, 2566 (1991); M.~Gasperini and M.~Giovannini, Phys.\ Lett.\ B {\bf 282}, 36 (1992).
 
\bibitem{thorne} K.S. Thorne, in: S. Hawking, W. Israel (Eds.) {\it 300 Years of Gravitation}, (Cambridge University Press, Cambridge, 1987).
 
 \bibitem{onea} D.~N.~Spergel {\it et al.},   Astrophys.\ J.\ Suppl.\  {\bf 148}, 175 (2003); D.~N.~Spergel {\it et al.},
{\em ibid.} \ {\bf 170}, 377 (2007);  L.~Page {\it et al.} {\em ibid.}  {\bf 170}, 335 (2007); B.~Gold {\it et al.},  Astrophys.\ J.\ Suppl.\ {\bf 192}, 15 (2011); 
D.~Larson,  {\it et al.},  {\em ibid.}  {\bf 192}, 16 (2011); C.~L.~Bennett {\it et al.}, {\em ibid.}\ {\bf 192}, 17 
(2011); G.~Hinshaw {\it et al.},  {\em ibid.} {\bf 208} 19 (2013); C.~L.~Bennett {\it et al.},   {\em ibid.} {\bf 208} 20 (2013); 
M.~Giovannini, Class.\ Quant.\ Grav.\  {\bf 31}, 225002 (2014);  P.~A.~R.~Ade {\it et al.}  [Planck Collaboration],   Astron.\ Astrophys.\  {\bf 571}, A22 (2014); 
Astron.\ Astrophys.\  {\bf 571}, A16 (2014);  P.~A.~R.~Ade {\it et al.} [Planck Collaboration], Astron.\ Astrophys.\  {\bf 594}, A20 (2016); M.~Giovannini, Phys.\ Lett.\ B {\bf 759}, 528 (2016).

\bibitem{oneb} G. Aad et al. [ATLAS Collaboration], Phys. Lett. B {\bf 716}, 1 (2012);  S. Chatrchyan et al. [CMS Collaboration], Phys. Lett. B {\bf 716}, 3 (2012).

\bibitem{two} M.~D'Onofrio and K.~Rummukainen, Phys.\ Rev.\ D {\bf 93}, no. 2, 025003 (2016);  K.~Kajantie, M.~Laine, J.~Peisa, K.~Rummukainen and M.~E.~Shaposhnikov,  Nucl.\ Phys.\ B {\bf 544}, 357 (1999); K.~Kajantie, M.~Laine, K.~Rummukainen and M.~E.~Shaposhnikov,
   Phys.\ Rev.\ Lett.\  {\bf 77}, 2887 (1996).

\bibitem{three} M.~Giovannini, Phys.\ Rev.\ D {\bf 61}, 063004 (2000);  Phys.\ Rev.\ D {\bf 61}, 063502 (2000).   
   
\bibitem{four} L.~Campanelli and M.~Giannotti, Phys.\ Rev.\ Lett.\  {\bf 96}, 161302 (2006); K.~Bamba, C.~Q.~Geng and S.~H.~Ho,  Phys.\ Lett.\ B {\bf 664}, 154 (2008);  L.~Campanelli, Int.\ J.\ Mod.\ Phys.\ D {\bf 18}, 1395 (2009); M.~Giovannini,  Phys.\ Rev.\ D {\bf 92}, no. 12, 121301 (2015); T.~Fujita and K.~Kamada,
   Phys.\ Rev.\ D {\bf 93}, no. 8, 083520 (2016);  A.~J.~Long and E.~Sabancilar,  JCAP {\bf 1605}, no. 05, 029 (2016); 
   K.~Kamada and A.~J.~Long,  Phys.\ Rev.\ D {\bf 94}, no. 12, 123509 (2016).

\bibitem{five} M.~Giovannini,  Phys.\ Rev.\ D {\bf 88}, 063536 (2013);  Phys.\ Rev.\ D {\bf 93}, no. 10, 103518 (2016); N.~Yamamoto,
  Phys.\ Rev.\ D {\bf 93}, no. 12, 125016 (2016); K.~Hattori and Y.~Yin,
  Phys.\ Rev.\ Lett.\  {\bf 117}, no. 15, 152002 (2016);  M.~Giovannini,  Phys.\ Rev.\ D {\bf 94}, no. 8, 081301 (2016).

\bibitem{fivea} V.A. Rubakov, Prog. Theor. Phys. {\bf 75}, 366 (1986).; V.A. Matveev {\it et al.}, Nucl. Phys. B {\bf 282}, 700 (1987); 
V.A. Rubakov, A.N. Tavkhelidze, Phys. Lett. B {\bf 165}, 109 (1985); A. N. Redlich and L. C. R. Wijewardhana, Phys. Rev. Lett. {\bf 54}, 970 (1984);
 D. Deryagin, D. Grigoriev, V. Rubakov, and M. Sazhin, Mod. Phys. Lett. A {\bf 11}, 593  (1986).
 
 \bibitem{seven} D. Kharzeev, L. McLerran and H. Warringa, Nucl. Phys. A {\bf 80}, 3227 (2008); 
K.~Fukushima, D.~Kharzeev and H.~Warringa, Phys.\ Rev.\ D {\bf 78}, 074033 (2008);
D.~Kharzeev,  Annals Phys.\  {\bf 325}, 205 (2010); S. Ozonder, Phys. Rev. C {\bf 81}, 062201 (2010); 
K.~Landsteiner, E.~Megias, L.~Melgar and F.~Pena-Benitez, JHEP {\bf 1109}, 121 (2011); Fortsch.\ Phys.\  {\bf 60}, 1064 (2012).

\bibitem{sevena}  M.~Giovannini and M.~E.~Shaposhnikov,  Phys.\ Rev.\ D {\bf 57}, 2186 (1998); Phys.\ Rev.\ Lett.\  {\bf 80}, 22 (1998).

\bibitem{sak} A.~D.~Sakharov, Pisma Zh.\ Eksp.\ Teor.\ Fiz.\  {\bf 5}, 32 (1967)  [JETP Lett.\  {\bf 5}, 24 (1967)].
  
\bibitem{CKN} A.~G.~Cohen, D.~B.~Kaplan and A.~E.~Nelson,  Ann.\ Rev.\ Nucl.\ Part.\ Sci.\  {\bf 43}, 27 (1993);
  W.~Buchmuller, R.~D.~Peccei and T.~Yanagida,  Ann.\ Rev.\ Nucl.\ Part.\ Sci.\  {\bf 55}, 311 (2005).

\bibitem{vain} A. P. Kazantsev, Zh. Eksp. Teor. Fiz. {\bf 53}, 1806 (1967) [Sov. Phys. JETP {\bf 26}, 1031 (1968)]; S. I. Vainshtein, Sov. Phys. Dokl. 15, 1090 (1971) [Dokl. Akad. Nauk SSSR 195, 793 (1970)]; Zh. Eksp. Teor. Fiz. 61, 612 (1971) [Sov. Phys. JETP 34, 327 (1971)]; S. I. Vainshtein and Ya. B. Zeldovich, Sov. Phys. Usp. {\bf 15}, 159 (1972) [Usp. Fiz. Nauk. {\bf 106}, 431 (1972)]. 

\bibitem{ax3} S. Carroll, G. Field and R. Jackiw, Phys. Rev. D {\bf 41},  1231 (1990);  W. D. Garretson, G. Field and S. Carroll, Phys. Rev. D {\bf 46}, 5346 (1992); G. Field and S. Carroll, Phys.Rev.D {\bf 62}, 103008 (2000).

\bibitem{such} G. Feinberg and J. Sucher, Phys. Rev. A {\bf 2}, 2395 (1970); Phys. Rev. D {\bf 20}, 1717 (1979). 

\bibitem{SUSC1} M.~Giovannini,  Phys.\ Rev.\ D {\bf 88}, 083533 (2013);  Phys.\ Rev.\ D {\bf 89},  063512 (2014); 
  Phys.\ Rev.\ D {\bf 92}, 043521 (2015).

\bibitem{six} E. Fermi and S. Chandrasekhar, Astrophys. J. {\bf 118}, 116 (1953);  S. Chandrasekhar and P. C. Kendall, Astrophys. J. {\bf 126}, 457 (1957); S. Chandrasekhar and L. Woltjer, Proc. Natl. Acad. Sci. U.S.A. {\bf 44}, 285 (1958); L. Woltjer, Proc. Natl. Acad. Sci. U.S.A. {\bf 44}, 489 (1958). 

\bibitem{sixa} R.~Jackiw and S.~-Y.~Pi, Phys.\ Rev.\ D {\bf 61}, 105015 (2000); C.~Adam, B.~Muratori and C.~Nash,
ÊÊPhys.\ Rev.\ D {\bf 61}, 105018 (2000); Phys.\ Rev.\ D {\bf 62}, 105027 (2000); Phys.\ Rev.\ D {\bf 66}, 103503 (2002).

\bibitem{ford} L. H. Ford and L. Parker, Phys. Rev. D {\bf 16},1601 (1977);  Phys. Rev. D {\bf 16}, 245  (1977); E.~D.~Schiappacasse and L.~H.~Ford,
  Phys.\ Rev.\ D {\bf 94}, no. 8, 084030 (2016).

\bibitem{sixb}   R. Isaacson, Phys. Rev. {\bf 166}, 1263 (1968); Phys. Rev. {\bf 166}, 1272 (1968);  L.~Abramo, Phys.\ Rev.\  D {\bf 60}, 064004 (1999); M.~Giovannini,  Phys.\ Rev.\  D {\bf 73}, 083505 (2006); Phys.\ Lett.\ B {\bf 668}, 44 (2008); Class.\ Quant.\ Grav.\  {\bf 26}, 045004 (2009).

 \bibitem{pulsar}  V.~M.~Kaspi, J.~H.~Taylor, and M.~F.~Ryba,   Astrophys.\ J.\ {\bf 428}, 713 (1994);
  F.~A.~Jenet {\it et al.}, Astrophys.\ J.\ {\bf 653}, 1571 (2006);  P.~B.~Demorest {\it et al.}, Astrophys.\ J.\  {\bf 762}, 94 (2013); R.~M.~Shannon {\it et al.}, 
  Science {\bf 349}, no. 6255, 1522 (2015); W. Zhao, Phys. Rev. D {\bf 83}, 104021 (2011); W. Zhao, Y. Zhang, X.-P. You, Z.-H. Zhu, Phys. Rev. D {\bf 87}, 124012 (2013). 

\bibitem{bbn} V.~F.~Schwartzmann, JETP Lett.\ {\bf 9}, 184
  (1969); M.~Giovannini, H.~Kurki-Suonio and E.~Sihvola, Phys.\ Rev.\  D {\bf 66}, 043504 (2002);
   R.~H.~Cyburt, B.~D.~Fields, K.~A.~Olive, and E.~Skillman, Astropart.\ Phys.\ {\bf 23}, 313 (2005).

\bibitem{space} S.~A.~Hughes,  Mon.\ Not.\ Roy.\ Astron.\ Soc.\  {\bf 331}, 805 (2002); arXiv:0711.0188 [gr-qc]; 
V.~Corbin and N.~J.~Cornish,  Class.\ Quant.\ Grav.\  {\bf 23}, 2435 (2006); S.~Kawamura {\it et al.}, J.\ Phys.\ Conf.\ Ser.\  {\bf 120}, 032004 (2008); 
A.~Nishizawa, E.~Berti, A.~Klein and A.~Sesana,Phys.\ Rev.\ D {\bf 94}, no. 6, 064020 (2016)

\bibitem{ligo} A.~Abramovici {\it et al.}, Science {\bf 256}, 325 (1992); B.~P.~Abbott {\it et al.} [LIGO/Virgo Collaboration],
 Phys.\ Rev.\ Lett.\  {\bf 116}, no. 6, 061102 (2016); B.~P.~Abbott {\it et al.} [LIGO/Virgo Collaboration],
 arXiv:1612.02029 [gr-qc].


\end{thebibliography}
\end{document}